\documentclass[twocolumn,preprintnumbers,superscriptaddress,
endnote,amsmath,amssymb,nofootinbib,aps]{revtex4}

\usepackage{graphicx}
\usepackage{slashed}
\usepackage{bbm}
\usepackage{footmisc}

\newcommand{\rig}{\rightarrow}

\newcommand{\be}{\begin{eqnarray*}}
\newcommand{\ee}{\end{eqnarray*}}
\newcommand{\gl}[1]{(\ref{#1})}

\newcommand{\bee}{\begin{eqnarray}}
\newcommand{\eee}{\end{eqnarray}}
\newcommand{\beeq}{\begin{equation}}
\newcommand{\eeeq}{\end{equation}}
\newcommand{\gev}{~{\text{GeV}}}
\newcommand{\br}{{\text{BR}}}
\newcommand{\ifb}{{\text{fb}}^{-1}}
\newcommand{\tev}{{\text{TeV}}}

\begin{document}

\title{Evasive Higgs Maneuvers at the LHC}

\begin{abstract}
  Non-standard decays of the Higgs boson produced at the Large Hadron
  Collider can lead to signatures which can easily be missed due to
  non-adapted trigger or search strategies.  Keeping electroweak symmetry breaking
  Standard Model-like we classify the phenomenology of an evasive
  Higgs boson into three categories and discuss how they can be
  described in an effective field theory. We comment on how one can
  improve the search strategies to also detect such an evasive Higgs.
\end{abstract}

\author{Christoph Englert} \email{christoph.englert@durham.ac.uk}
\affiliation{Institut f\"ur Theoretische Physik, Universit\"at Heidelberg, 69120
  Heidelberg, Germany}
\affiliation{Institute for Particle Physics Phenomenology, Department
  of Physics,\\Durham University, DH1 3LE, United Kingdom}
\author{Joerg Jaeckel} \email{joerg.jaeckel@durham.ac.uk}
\affiliation{Institute for Particle Physics Phenomenology, Department
  of Physics,\\Durham University, DH1 3LE, United Kingdom}
\author{Emanuele Re} \email{emanuele.re@durham.ac.uk}
\affiliation{Institute for Particle Physics Phenomenology, Department
  of Physics,\\Durham University, DH1 3LE, United Kingdom}
\author{Michael Spannowsky} \email{michael.spannowsky@durham.ac.uk}
\affiliation{Institute for Particle Physics Phenomenology, Department
  of Physics,\\Durham University, DH1 3LE, United Kingdom}

\pacs{}
\preprint{IPPP/11/73\;\;\;\;DCPT/11/146\;\;\;\;LPN11-61}

\maketitle

\section{Introduction}

One of the main goals of the Large Hadron Collider (LHC) is the
discovery of the Higgs boson responsible for electroweak symmetry
breaking \cite{searches}. With the LHC having provided about $5~\ifb$
of data, both {\sc{Atlas}} and {\sc{Cms}} have presented strong
constraints on a Standard Model (SM) Higgs boson. If the Higgs is
SM-like it is disfavored at the 95\% confidence level in the mass
range 131~GeV~(127~GeV) to 453~GeV~(600~GeV) by
{\sc{Atlas}}~({\sc{Cms}}) \cite{ATLASnew,CMSnew}, while the LEP2 bound
\cite{LEP2} of 114.4~GeV is pushed to 115.5~GeV~(115~GeV).  With the
rapidly growing data sample we can expect to have full coverage of the
SM Higgs mass at the 95\% confidence level all the way down to 115~GeV
at the LHC for an integrated luminosity of $6~\ifb$ \cite{atlasprosp}.

The ${\cal{O}}(3\sigma)$ excesses measured by both {\sc{Atlas}}
\cite{atlashiggs,ATLASnew} and {\sc{Cms}} \cite{cmshiggs,CMSnew} in
the $h\to \gamma \gamma$ channel might be the first glimpse of a light
Higgs (at a considerably larger than expected production cross
section) which would be in perfect agreement with all we expect from
electroweak precision constraints and measurements performed at LEP
and at the Tevatron \cite{lepteva}. However, this might also be just a
fluctuation and the Higgs may yet again escape.

What are the implications of these first couple of inverse femtobarns
of data for more involved (and better motivated) scenarios of symmetry
breaking? {\sc{Atlas}} and {\sc{Cms}} have also presented stringent
bounds on new physics such as, e.g., the minimal supersymmetric
extension of the SM and the SM with the addition of fourth
generation. And we expect more model-specific analyses to appear soon,
when more data becomes available.

Both {\sc{Atlas}} and {\sc{Cms}} provide strong constraints on SM-like
Higgs production, even if it is suppressed compared to the SM.  These
constraints are expressed as limits on $\sigma/\sigma_{\text{SM}}$ and
in some mass ranges $\sigma/\sigma_{\text{SM}}\simeq 0.2$ is already
excluded at the 95\% confidence
level~\hbox{\cite{atlashiggs,cmshiggs,ATLASnew,CMSnew,combination}}.
If a Higgs is to exist in this energy range it has to somehow ``hide''
from the Standard searches.

The relevant features for a Higgs search are the production rate and
the decay signatures in the detector. Therefore one option to hide the
Higgs is to try and significantly reduce the production of the Higgs.
Alternatively, the Higgs could dominantly decay into particles which
are difficult to detect or difficult to disentangle from the
background. 

Constraints on our ability to reduce the production cross section
arise from the fact that Higgs production rates in the SM are bound to
the physical Higgs being the unitarizing degree of freedom in
longitudinal gauge boson scattering $V_LV_L\to V_LV_L$ ($V=W^\pm, Z$)
and in massive $q\bar q \to V_L V_L$ amplitudes\footnote{Unitarization
  of $V_LV_L\to V_LV_L$ is a direct consequence of spontaneous
  symmetry breaking \cite{cornwall}, while unitarity in $q\bar q \to
  V_L V_L$ relates the fermion and gauge sectors and is less obvious,
  see e.g. Ref.~\cite{Hagiwara:1986vm}.}. In the SM this fixes the
partial decay widths and the production cross section for associated
Higgs production and weak boson fusion.  Essentially this limits the
total Higgs production cross section from below (interference effects
are typically small~\cite{Bredenstein:2008tm,Ciccolini:2007jr}).
Gluon fusion, $gg\to h$, is sensitive to the propagating heavy
fermionic degrees of freedom and the production rate is again fixed as
a function of the Higgs-fermion couplings.  Any new physics extension
or modification of the Higgs sector has to reproduce these unitarity
restoring features to leave a theoretically sound and predictive
theory in the LHC-accessible energy range of $\lesssim 3$ TeV in the
weak boson fusion channel \cite{wbf}, which directly accesses
longitudinal gauge boson scattering. In extended Higgs sectors this is
typically achieved by linear mixing of the various scalar fields $h_i$
such that for energies much larger than $m_{h_i}$ the coherent sum of
the $h_i$ exchange diagrams reproduces the SM Higgs contribution. This
is the case for e.g. the two Higgs doublet model, the next to minimal
supersymmetric Standard Model~\cite{Ellwanger}, composite Higgs models
\cite{Gripaios:2009pe} or the Higgs portal scenarios of
Refs.~\cite{Strassler:2006im,portal}. As a result the Higgs production
rates (or more precisely the production rate of the light SM-like
Higgs) can decrease with a characteristic mixing angle. This can also
protect precision electroweak observables such as $S,T,U$
\cite{peskinstu} from sizable corrections from high mass scales.
However, unless we appeal to fine-tuning, accounting for both
electroweak precision data and unitarity at the same time typically
amounts to a light SM-like Higgs boson with significant production
cross section~\cite{ewpt,Englert:2011yb}. Finally, smaller Higgs
production cross sections can also be obtained from anomalous Higgs
couplings \cite{anomalous}.

The most minimal assumption is that electroweak symmetry breaking is
caused by a single $SU(2)$ doublet Higgs.  In this paper we will use
this assumption and therefore we will focus on invisible decays and
modified signatures.  In the following we investigate how a simple
extension of the Higgs sector can lead to a ``hidden'' Higgs
phenomenology at the LHC. Generically, the recent LHC bounds can be
weakened or even avoided this way.

Such Higgs ``hide-out'' scenarios are due to a combination of dynamics
and kinematics, e.g. they occur through modified branching ratios as a
consequence of an extended spectrum or modified couplings.  There is a
plethora of theoretically sound models which do such modifications and
hide the
Higgs~\cite{portal,Goldberger:2007zk,Foot:2011xh,Englert:2011yb,Bellazzini:2009xt}.
Therefore, a classification on the level of the phenomenological
outcome is desirable.  In the following we will pursue this approach.
Nevertheless, we also provide an interpretation in terms of a simple
effective model.  Our categorization is of course ``non-invertible'';
many different models \cite{Alves:2011wf} exhibit a similar
phenomenology, and we do not try to compile an exhaustive list of
model-building realizations of a specific phenomenological outcome.  A
minimal realization is the coupling of the otherwise SM-like Higgs to
a hidden sector via a renormalizable ``portal''
interaction\footnote{There are two more such portals: kinetic mixing
  with an extra $U(1)$ gauge group (see e.g.~Ref.~\cite{Holdom:1985ag}
  and Ref.~\cite{Jaeckel:2010ni} for a review of some low energy
  consequences), and neutrinos mixing with sterile neutrinos
  (cf.~\cite{neutrinos}).}  $\sim |H|^2{\cal{O}}_{hid}$
\cite{portal}. Although it is one of the simplest gauge invariant and
renormalizable extensions we can come up with to model a Higgs
hide-out, it is an example of how to evade the currently existing
bounds on SM Higgs and encompasses a huge range of phenomenological
characteristics.

Starting from our assumption of an essentially SM-like electroweak
symmetry breaking mechanism, our aim is to categorize the variety of
Higgs hide-outs in terms of their phenomenological signatures. At the
same time we want to provide a simple parametrization in terms of an
effective Lagrangian that realizes these features.  In section
\ref{sec:pheno} we review and collect the necessary ingredients for
the Lagrangian.  In Sect.~\ref{subsec:1} we consider a dominantly
invisibly decaying Higgs and study the implications of current LHC
data.  Then in Sect.~\ref{subsec:2a} we look at a Higgs decaying into
long lived particles that could be searched for by displaced vertex
searches.  Here we provide additional motivation to also search in the
outer parts of the detector.  Finally the Higgs could also be buried
in large SM backgrounds (Sect.~\ref{subsec:2}).  While this cannot be
achieved by changing the gluon coupling alone we find that it is
possible in scenarios with enhanced couplings to light quarks or when
heavy flavor mesons decaying into gluons dominate the Higgs decay
chain.  We briefly discuss combinations of these scenarios in
Sect.~\ref{subsec:3}.  We conclude this paper with a summary in
section \ref{sec:summary}.

\section{Phenomenological Higgs hide-outs}
\label{sec:pheno}
To achieve a situation in which the Higgs phenomenology can be hidden
we have to introduce an extension of the SM Higgs sector which
preserves $SU(2) \times U(1)$ gauge invariance. We limit ourselves to
renormalizable interactions in the Higgs sector. The only choice is
the addition of a scalar interacting via the previously mentioned
Higgs portal
\begin{equation}
  \begin{split}
  \label{eq:portal}
  {\cal{L}}=&{\cal{L}}_{\text{SM}}+\eta |H|^2 |\phi|^2 + \partial_\mu
  \phi^\star \partial^\mu\phi
  - m^2 |\phi|^2. \\
\end{split}
\end{equation}

The field $\phi$ is taken to be a singlet under the SM gauge
group. One could have non-trivial representations of $\phi$ under
$SU(2)_L\times U(1)_Y$ in the Higgs sector, which admits more involved
dynamics than Eq.~\gl{eq:portal}.  If these extra degrees of freedom
are phenomenologically hidden, we encounter a situation which still
can be meaningfully described by Eq.~\gl{eq:portal}.  Throughout, we
define the field $H$ to be responsible for electroweak symmetry
breaking. $\phi$ can live in the scalar or vectorial representation of
the Lorentz group but can in principle also effectively arise from a
fermionic condensate of a strongly interacting sector.  As an
additional simplification one can impose a $U(1)$ or ${\mathbbm{Z}}_2$
symmetry which forbids terms $\sim \phi$.

In general there could also be more than one field $\phi$. This can
allow for a variety of decay cascades in the hidden sector.  In our
study we will concentrate on the following simple set of interactions,
\begin{eqnarray}
  \label{eq:cascades}
  \nonumber
  {\cal{L}}_{\text{multi}}\!\!&=&\!\!\sum^{n}_{i=1} \partial_{\mu}\phi^{\star}_{i}\partial^{\mu}
  \phi_{i}-m^2_{i}\phi^{\star}_{i}\phi_{i}+\partial_{\mu}\phi^{\prime\star}_{i}
  \partial^{\mu}\phi^{\prime}_{i}-m^{\prime 2}_{i}\phi^{\prime \star}_{i}\phi^{\prime}_{i}
  \\
  \!\!&+&\!\!\sum^{n-1}_{i=1}\rho_{i}\phi^{\star}_{i}\phi_{i+1}\phi^{\prime}_{i+1}
   +{\text{h.c.}}\,.
\end{eqnarray}
If desired one can choose charges such that the $U(1)$ symmetry is
preserved.  We take the particles to be ordered in mass, allowing
cascade decays. We identify the heaviest state $\phi_{1}$ with $\phi$
in Eq.~\eqref{eq:portal}.  Also we will typically allow only the last
particles $\phi_{n}$ of the decay cascade to decay into SM particles
accordingly. If a cascade is considered one has to replace $\phi\to
\phi_{n}$ in Eq.~\eqref{violcouplings} below.

After the Higgs acquires a vacuum expectation value we induce a BSM
trilinear coupling of the physical Higgs $h$ to $\phi$
\begin{equation}
  \epsilon h \phi^{\star}\phi \quad{\text{with}}\quad 
  \epsilon= \eta \left\langle H\right\rangle=\eta v/\sqrt{2}
\end{equation} 
which modifies the Higgs branching ratios for the Higgs decay.  In
addition the physical mass $m_{\phi}$ is given by
\begin{equation} 
m^{2}_{\phi}=m^{2}-\eta\langle H\rangle^2=m^{2}-\eta v^2/2,
\end{equation}
which we take to be positive.
Note that since $\phi$ does not develop a vacuum expectation value,
there is no mixing of the two scalar states. Accordingly cross
sections and decay rates are not modified by mixing effects.

If unbroken, the global $U(1)$ symmetry forbids decays of $\phi$ into
SM particles.  Hence, in order to re-introduce such decays we need
$U(1)$ breaking couplings to SM particles.  The SM gauge symmetries
forbid couplings of SM matter and gauge fields to $\phi$ on the
renormalizable level. We will therefore consider the following
dimension 5 couplings (which automatically also explicitly break the
$U(1)$ symmetry),
\begin{multline}
  \label{violcouplings}
  {\cal{L}}\supset \sum_{i} \frac{\lambda_{ij}}{M}\phi \overline
  D_{L,i}H \Psi_{R,j} +
  {\text{~h.c.~}} \\
  +\left[\frac{\kappa_{\gamma}}{M}(\phi+\phi^{\star})F^{\mu\nu}F_{\mu\nu}+
    \frac{\tilde{\kappa}_{\gamma}}{M}(\phi-\phi^{\star})
    F^{\mu\nu}\tilde{F}_{\mu\nu}\right] \\
  +\left[\frac{\kappa_{g}}{M}(\phi+\phi^{\star})G^{\mu\nu}G_{\mu\nu}+
    \frac{\tilde{\kappa}_{g}}{M}(\phi-\phi^{\star})
    G^{\mu\nu}\tilde{G}_{\mu\nu}\right] \,
\end{multline}
where $D_L$ denotes the left-handed fermion doublet and $\Psi_R$ the
right handed fermions.  In our effective theory approach we can choose
suitable $\lambda_{ij}$ to avoid flavor, lepton number and baryon
number changing processes~\cite{Froggatt:1978nt}.  $\tilde F,~\tilde
G$ are the dual QED and QCD field strength tensor\footnote{If
  kinematically possible, i.e. for very high Higgs masses, one could
  also add an additional term for decays to $W^{\pm},Z$.  This however
  would not lead to a phenomenologically very different situation.}.
These higher dimensional operators allow $\phi$ to decay back to
visible SM matter.

Since we take the Higgs to be responsible for electroweak symmetry
breaking the partial decay widths for $h\to VV$ are fixed.  However, we
can model modifications of the branching ratios and the total decay
width by introducing an additional contribution to the coupling
\begin{equation} 
  \label{eq:gghiggs}
  {\cal L}_{ggh}=\chi\frac{\alpha_{s}}{12\pi v}h
  G^{\mu\nu}G_{\mu\nu}\, ,
\end{equation}
(see Ref.~\cite{Low:2009di} for theoretical bounds) or operators of
the form
\begin{equation}
  \label{moddecay}
  {\cal L}_{HHqq}=\frac{\beta_{Q_{L}}}{M^2}H^{\dagger}
  H\bar{Q}_{L}\slashed{D}Q_{L}.
\end{equation}
Of course one can also add similar couplings for the right handed
quarks and leptons.

\subsection{Hidden Higgses}
\label{subsec:1}
If the global $U(1)$ symmetry of the $\phi$-field is unbroken or
extremely weakly broken, i.e. the couplings given in
Eq.~\eqref{violcouplings} vanish or are very small, $\phi$ is stable
with respect to decays into Standard Model particles. Accordingly, the
decay $h\to\phi^{\star}\phi$ (possible as long as $m_{h}\geq
2m_{\phi}$) is invisible.

Such invisible decays make search strategies based on visible SM
particles more difficult.  Naively, one can again use that current
Higgs searches are already sensitive to cross sections lower than the
SM cross section.  Using this one can reinterpret the limits on
$\sigma_h/\sigma_h^{\text{SM}}$ as limits on $\sigma_h \br(h\to
{\text{SM}})/\sigma_h^{\text{SM}}$.

The partial decay width for $h\to\phi^{\star}\phi$ is given by
\begin{equation}
  \Gamma_{\text{inv}}= {\eta^2v^2\over 16\pi} 
  { [(m_h^2-4m_\phi^2) ]^{1/2} \over m_h^2}.
\end{equation}

\begin{figure}[t]
  \centering
  \hspace{-0.3cm}
  \includegraphics[height=5.5cm]{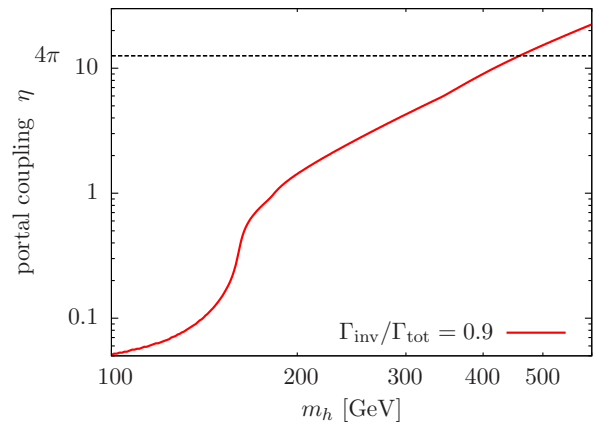}
  \caption{\label{fig:etacrit1} Required size of the Higgs portal
    coupling $\eta$ to achieve an invisible branching ratio of
    $\br({\text{invis}})=90\%$. We choose $m_\phi=10\gev$, the result
    is however rather independent of this choice.}
\end{figure}

As can be seen from Fig.~\ref{fig:etacrit1} at low Higgs masses quite
moderate values of $\eta$ are sufficient to achieve dominantly
invisible decays, allowing to evade search strategies based on SM
particles for now. Even when we later on look at modified decays to SM
particles this makes it more difficult to hide the Higgs in the high
mass region.  Above the $VV$-threshold, however, the hidden sector
decay width has to compete with a much larger and rapidly growing
$\sim m^{3}_{h}$ decay width into SM particles. This requires fairly
large values of the Higgs portal coupling $\eta\gtrsim 1$ for
$m_{h}\gtrsim 180\gev$ and even $\eta\gtrsim 4\pi$ for $m_{h}\gtrsim
450\gev$. The reason for this is as simple as compelling:
Unitarization of $VV$ scattering requires a sufficiently large
coupling of the Higgs to energetic longitudinal $V$s resulting also in
a large particle decay width of the Higgs into those if the Higgs is
heavy.  The required large couplings to hide the Higgs into invisible
decays are at odds with perturbativity unless they are connected to
spontaneous symmetry breaking. Therefore, at large Higgs masses hiding
the Higgs into invisible decays is difficult from a model building
point of view. For low Higgs masses the Higgs can be easily hidden in
invisible decays and alternative strategies \cite{invisible,Belotsky:2004ex} 
based on
missing energy ($\slashed{E}_T$) searches then become necessary.

An important constraint for hidden Higgses comes from direct searches
of associated hidden Higgs production at LEP.  For SM-like Higgs gauge
boson couplings this puts a lower limit of $114.4\gev$ on the mass of
a dominantly invisibly decaying Higgs~\cite{:2001xz}.

\subsubsection*{Monojet + missing energy channel}

At the LHC the production of a light Higgs boson is dominated by the
gluon fusion process $gg\rig h+X$. Accordingly a search in the monojet
plus missing energy channel is promising \cite{Birkedal:2004xn}
because backgrounds are comparably small and can be brought under
sufficient control~\cite{Atlasinv}.  We focus in the following on a
center of mass energy of $7~{\text{TeV}}$, which allows to relate our
results to the current LHC run.

We compute the $gg\rightarrow h+X$ signal using an NLO computation
matched to a parton shower\footnote{The matching prescription in this
  channel is subject to an ongoing discussion in the corresponding
  community, see Refs.~\cite{Hoeche:2011fd,powheg} for details. Our
  results do not include any theoretical uncertainties.}: the
inclusive cross section is therefore NLO accurate, the hardest jet
(relevant for this study) is described with the full $h+1$ jet matrix
element accuracy and further emissions are generated in the shower
approximation. We use the {\sc{Powheg}} method (as implemented in the
{\sc Powheg Box} program) \cite{powheg}, together with
(transverse-momentum ordered) {\sc{Pythia}}~6~\cite{pythia6}. We
generate events for values of $m_h$ between $100$ and $500$ GeV, using
the narrow-width approximation.
 
\begin{figure}[t]
  \centering
  \includegraphics[height=5.5cm]{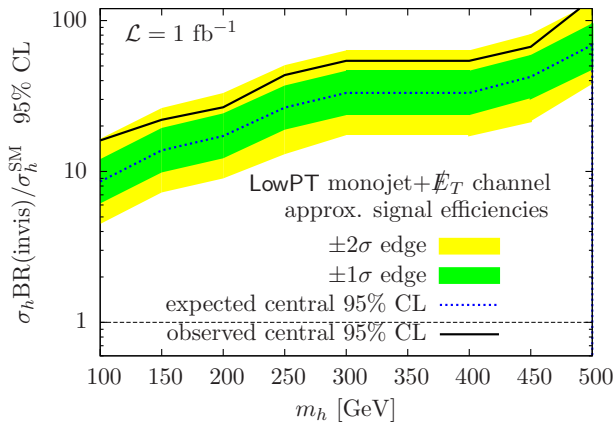}
  \caption{\label{fig:cls1} Expected and observed
    ($\left\langle\xi\right\rangle=0.27$) 95\% CL upper limits on the
    production cross section in multiples of the SM cross section for
    a monojet+$\slashed{E}_T$ search confronted with an invisibly
    decaying Higgs. }
\end{figure}

The Higgs-boson is set stable and excluded from the tracks entering
the analysis. On all the other final state particles we apply the
``{\sf{LowPT}''} selection cuts described in
Ref.~\cite{Atlasinv}. Jets are constructed using the anti-$k_t$
algorithm~\cite{Cacciari:2008gp}, with $R=0.4$, and we keep only
events where
\begin{eqnarray}
\label{monocut}
&& p_T^{j_1} > 120\ \mbox{GeV},\ |\eta^{j_1}| < 2 \nonumber\\
&& p_T^{j_2} < 30\ \mbox{GeV} \ \ \mbox{if}\ |\eta^{j_2}| < 4.5 \ (\mbox{jet-veto}) \nonumber\\
&& \slashed{E}_T > 120\ \mbox{GeV}\,.
\end{eqnarray}
We obtain $\slashed{E}_T$ as the transverse momentum of the total
momentum from all the visible particles, defined to be final state
particles with $p_T > 2$ GeV and $|\eta| < 4.5$. We also rejected the
(rare) events where there is at least one lepton within the cuts
defined in Ref.~\cite{Atlasinv}.

We extract the background distribution and the data for a luminosity
${\cal{L}}=1~\ifb$ from Ref.~\cite{Atlasinv}.

In Fig.~\ref{fig:cls1} we plot the resulting 95\% confidence level
(CL) upper bound on
$\sigma_{h}\br({\text{invis}})/\sigma_h^{{\text{SM}}}$
applying the CL$_S$ method\footnote{The CL$_S$ method procedure is
  fairly standard method in nowadays experimental analysis to present
  constraints in new physics searches. We refer the read to
  Refs.~\cite{Read:2002hq,junk} for further
  details.}. 
  
We estimate the average efficiency $\left\langle \xi \right \rangle$
that relates the theoretical Higgs cross section prediction for a
given mass to the experimentally observable one by performing a Monte
Carlo study relating the results of the dominant and signature-wise
similar $(Z\to{\hbox{inv}})$+jets background to the expected numbers
quoted in Ref.~\cite{Atlasinv}. For this purpose we produce a
$(Z\to{\hbox{inv}})$+jets event sample using {\sc{Sherpa}}
\cite{Gleisberg:2008ta} and normalize it to the next-to-leading order
QCD cross section obtained with {\sc{Mcfm}}~\cite{mcfm}
($\sigma_{(Z\to \bar\nu \nu)+j}^{\text{NLO}}(p_T^j\geq
120\gev,|\eta_j| \leq 2.0)=41.3~{\text{pb}}$).  Running the analysis
with the above cuts we can estimate $\langle \xi \rangle=0.27$ and we
use this significance to rescale our Higgs signal hypotheses in
Fig.~\ref{fig:cls1}. The observed limits are weaker than the expected
limits because of a slight excess of the central data values
\cite{Atlasinv}.

The confidence level scales as $\sim{\cal{L}}^{-1/2}$. While searches
with present luminosities are insensitive to SM-like production cross
sections ($\chi \simeq 1$ in Eq.~\gl{eq:gghiggs}) enhanced production
cross sections ($\chi \simeq 3-5$) occurring in a variety of models
are already constrained. In particular for fourth generation models
\cite{4gen,Holdom:2009rf,Ruan:2011qg,Carpenter:2011wb} where we expect
$\chi\simeq 3$, this will very soon become a relevant constraint. With
the fast growing data sets these constraints will tighten
significantly in the near future.

\subsubsection*{Two Leptons + $\slashed{E}_T$ Channel}
A very clean and therefore important search channel for hidden Higgs
decays is associated production $pp\rig hZ$ with subsequent decay of
the $Z$ to leptons \cite{invisible,Davoudiasl:2004aj}. Note that for
associated production, it is much more difficult to obtain increased
cross sections since the relevant coupling is fixed by gauge
invariance\footnote{This fact also accounts for the hidden higgs
  search in the weak boson fusion channel discussed in
  \cite{Eboli:2000ze}. For a center of mass energy of 7 TeV weak boson
  fusion is not an important channel since for typical search cuts and
  ${\cal{L}}\simeq 1\ifb$. We find a reduced cross section
  \cite{vbfnlo} $\sigma(7~\tev)/\sigma(14~\tev) \simeq 0.2$ compared
  to 14~TeV. As a consequence searches based on small angular
  separations of the two tagging jets are less sensitive compared to
  other channels if one also takes into account the systematic
  uncertainties of the central jet veto and the forward tag jet energy
  scale uncertainty due to pile-up. Only recently, with the $5~\ifb$
  set {\sc{Cms}} has started to overcome these systematic limitations
  \cite{CMSwbf}.}.  For completeness and comparison with the previous
section, we nonetheless estimate the performance of the corresponding
search. To our knowledge there is no publicly available LHC result of
the processes in the phase space region we are interested in. Hence
the results of this section are obtained from Monte Carlo, and we
include neither background nor signal systematics (this includes a
potential mismeasurement of $Z$+jets, giving rise to a finite
$\slashed{E}_T$ by detector effects). Therefore, the sensitivity in
this channel is obviously optimistic.

The cross section for associated Higgs production is much smaller than
the one for gluon fusion, and shape comparisons of e.g. the $p_T^Z$
distribution are not possible given $\lesssim 10$ signal events for
${\cal{L}}={1~\ifb}$.  Instead we perform a counting experiment for
the signal and dominant $ZZ$, $WW$ and $t\bar t$ backgrounds. Again we
compute the signal and background cross sections at $\sqrt{s}=7~\tev$
using {\sc{Mcfm}}. We require two oppositely charged leptons of
identical flavor to combine to the $Z$ mass within a $\pm 10\gev$ mass
window and $\slashed{E}_T\geq 100\gev$.

\begin{figure}[t]
  \centering
  \includegraphics[height=5.5cm]{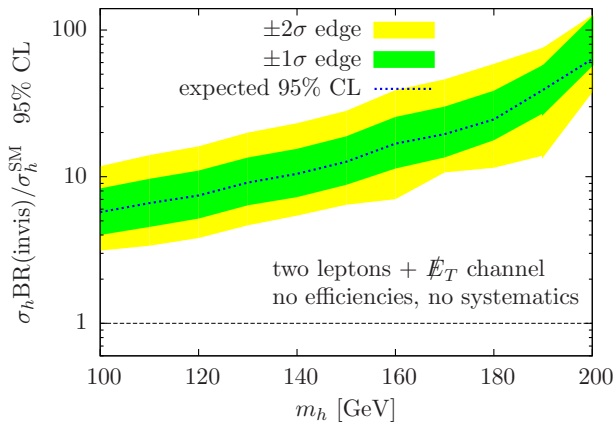}
  \caption{\label{fig:cls2} Expected 95\% CL upper limits on the
    associated production cross section in multiples of the SM cross
    section for a two leptons+$\slashed{E}_T$ search confronted with an
    invisibly decaying Higgs. Superficially this channel looks more
    sensitive then the monojet search. Note, however, that this plot
    does not include any efficiencies and is not based on actual
    data.}
\end{figure}

Our resulting estimate on the upper 95\% CL is shown in
Fig.~\ref{fig:cls2}. For very low masses the two
leptons+$\slashed{E}_T$ channel can be of similar importance as the
monojet+$\slashed{E}_T$ search, depending on the signal and background
efficiencies. For higher Higgs masses this channel looses sensitivity
very quickly due to the small cross section at 7~TeV center of mass
energy.  Eventually this channel will become again important for large
luminosities ${\cal{O}}(100~\ifb)$ at
$\sqrt{s}=14~\tev$~\cite{invisible}.

\begin{figure}[b]
\includegraphics[height=5.3cm]{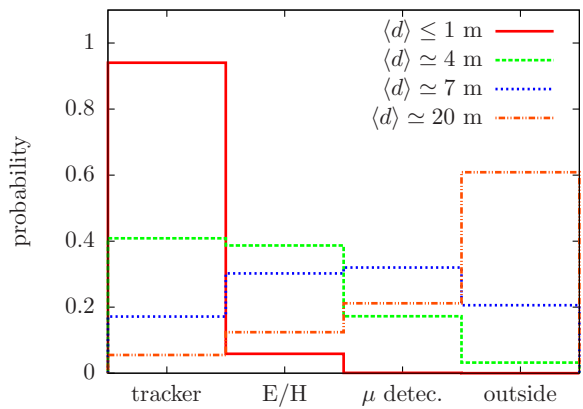}
\caption{\label{fig:reemerge}Probability, based on the {\sc{Atlas}}
  geometry, for a $\phi$ to decay in the tracker, E/H-calorimeter,
  muon-calorimeter or outside the detector. We show the results for
  four different average lab frame decay lengths, respectively.}
\end{figure}

\subsection{Reemerging Higgses}\label{subsec:2a}
In some cases the ``invisible'' decay products of the previous
subsection can decay back into SM particles and the hidden Higgs
slowly reemerges. In our toy model this is realized when the $U(1)$
violating couplings of Eq.~\eqref{violcouplings} are turned on. At
leading order the decay rate of $\phi$ is given by,
\begin{equation}
  \Gamma_{\phi}={Y^2 \over 4\pi}  {(m_\phi^2-4m_f^2)^{3/2} \over 2 m_\phi^2} 
\end{equation}
where $ Y={\lambda}v/({\sqrt{2}}{M})$ is the effective Yukawa coupling
that results from the lagrangian Eq.~\eqref{violcouplings}.  If the
couplings are very small, i.e. the mass scale $M$ is very high, $\phi$
can travel a measurable distance before decaying,
\begin{equation}
  d=\frac{\beta \gamma}{\Gamma_{\phi}}\,.
\end{equation}
In this case one may search for (highly) displaced vertices
\cite{Strassler:2006ri} or use adapted trigger strategies
\cite{trigger,theory}.  If decays happen inside the tracker this is a
fairly clean signature.  If the decay length is of the order of meters
and above a significant part of the decays will happen in the bigger
outer parts of the detector (or even outside the detector).  In this
case one can gain additional sensitivity by also triggering on events
where the decay happens outside of the tracker. If the decay happens
outside the detector coverage the same search strategies outlined in
Sec.~\ref{subsec:1} apply.

The LHC experiments {\sc{Atlas}} and {\sc{Cms}} consist each of four
different layers: the inner tracker, the electromagnetic (E)
calorimeter, the hadron (H) calorimeter and the muon calorimeter. The
sensitivity on highly displaced vertices is limited by the radial
extension of the experiments. The radial size of the different
detector components differs between the two experiments, particularly
for the muon calorimeters. We focus in the following analysis
optimistically on the geometry of the larger {\sc{Atlas}} experiment.

The muon calorimeter of the {\sc{Atlas}} experiment is located
$d\lesssim 11~{\text{m}}$ away from the interaction point. The muon
detector coverage in pseudorapidity is $|\eta|<2.4$. In the muon
calorimeter photons and electrons will be stripped in an early layer
after the conversion \hbox{$\phi\to f \bar f$}, hence, are likely to
be misinterpreted as detector noise~\cite{Ploch}. Assuming the Higgs
boson decays instantaneously the probability for $\phi$ to decay
between distances $d_1$ and $d_2>d_1$ is given by
\begin{equation}
  \label{eq:prob}
  p(d_1\leq d \leq d_2) = \int_{d_1}^{d_2} {\text{d}} x~{\Gamma\over\beta
    \gamma}   \exp\left({-\frac{\Gamma x}{\beta \gamma}} \right)\,.
\end{equation} 

In the lab frame the decay of $h\rightarrow \phi^\star \phi$ induces a
significant boost factor $\beta$ for $\phi$. We find, that the
transverse momentum of the Higgs boson generated from initial state
radiation is of minor importance for $\beta$, because the Higgs rarely
decays along its direction of motion.  Based on Eq.~(\ref{eq:prob}) we
show the probability of $\phi$ decaying respectively in the tracker,
the E/H-calorimeter, the muon-calorimeter or outside the detector in
Fig.~\ref{fig:reemerge} for a number of decay lengths. We include the
effects of finite detector coverage $|\eta |\leq 2.4$ for a particle
cluster with $p_T\geq 2~\gev$ and correct on the longitudinal
dimension of the detector.

\begin{figure}[t]
  \centering
  \includegraphics[height=5.2cm]{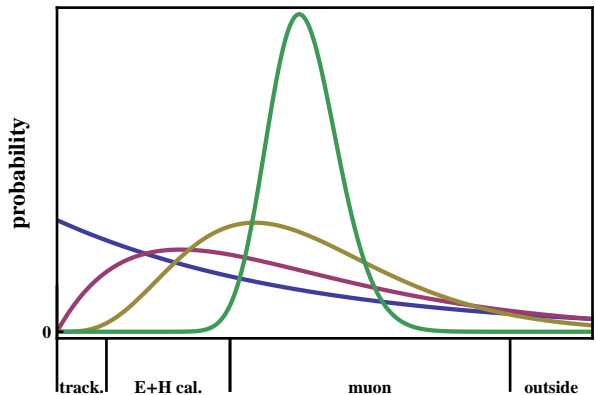}
  \caption{\label{fig:cascade} Probability for the distance between
    initial Higgs production and final decay back into SM particles
    for a cascade decay with 1 (blue), 2 (red), 5 (yellow) and 50
    (green) intermediate steps.}
\end{figure}

So far the larger number of decays happening in the outer parts of the
detector arise simply from the fact that these parts are bigger.
However, if we allow for cascade decays $h\to
\phi_1\to\phi_2\to\ldots\to {\text{SM}}$ as described by the
Lagrangian Eq.~\eqref{eq:cascades} one can achieve that more decays
happen close to the average total decay length.  For the simple case
when subsequent decays generate negligible transverse momentum
(i.e. when the mass difference to the sum of the masses of the decay
products is small) and the decay lengths are all equal $\lambda/n$ the
resulting probability distribution for an $n$-step decay is
\begin{equation}
  P_{n}(x)=\frac{\exp\left(-{nx}/{\lambda}\right) n (nx)^{n-1}\lambda^{-n}}{(n-1)!}\,.
\end{equation}
This is shown in Fig.~\ref{fig:cascade}.  In this way one can reduce
the number of decays happening in the inner tracker compared to decays
occurring in outer parts increasing the need to also check for decays
there.

\subsection{Buried Higgses}
\label{subsec:2}
Naively we can also try to hide the Higgs by burying it in the busy
hadronic final state at the LHC. In our effective theory approach we
can facilitate this by increasing its branching ratio to gluons via
the operator of Eq.~\gl{eq:gghiggs}. The modified decay width to
gluons then scales like
\begin{equation}
  \Gamma(h\to gg) = \chi^2\Gamma_{\text{SM}}(h\to gg)\,.
\end{equation}
The same effect increases the higgs production cross section from
gluon fusion by the same factor, yielding modified production times
branching ratio factors for the individual higgs decay channels $gg
\rig h \rig i i^{(\dagger)}$
\begin{equation}
\begin{split}
  i\neq g: \quad& {[\sigma\br] \over [\sigma\br]_{\text{SM}}} =
  {\chi^2 \over (\chi^2-1) \br_{\text{SM}}(h\rig g g) +1}\,,\\
  i= g: \quad& {[\sigma\br] \over [\sigma\br]_{\text{SM}}} = {\chi^4
    \over (\chi^2-1) \br_{\text{SM}}(h\rig g g) +1} \,.
\end{split}
\end{equation}
For the ``standard'' search channels we have
\begin{equation}
  {[\sigma\br] \over [\sigma\br]_{\text{SM}}} \gg 1
  \hbox{~for~} \chi \gg 1,
\end{equation}
since for light Higgs masses we have $\br_{\text{SM}}(h\to gg) =
{{\cal{O}}(1\%)}$.  As a result the Higgs is not buried but even more
visible in the standard search channels. Note, however, that the
increased total width of the Higgs boson will give rise to reduced
reconstruction efficiencies in standard search channels\footnote{In
  principle a similar effect can arise if we dramatically increase an
  invisible decay width. The monojet and other invisible Higgs search
  strategies should be less affected by this.}.

\begin{figure}[t!]
  \centering
  \includegraphics[height=5.5cm]{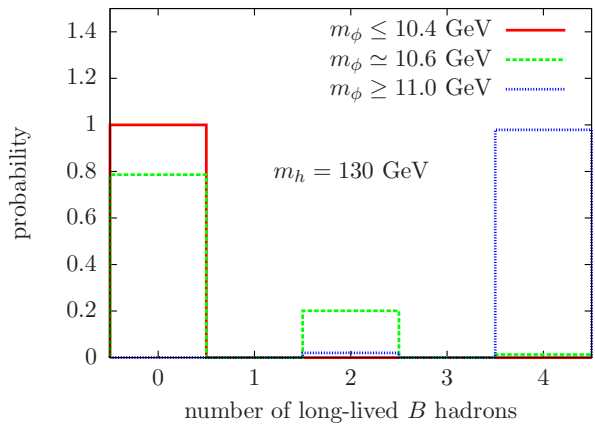}
  \caption{\label{fig:buried} We count the number of long lived
    $B$-hadrons with decay length $c\tau\geq 0.3~{\text{mm}}$ for a
    higgs decaying via $h\to \phi\phi\to
    b\bar{b}b\bar{b}$.}
\end{figure}

One way to bury the Higgs in hadronic final states is to enhance one
of the couplings to light quarks, for example using the operator of
Eq.~\eqref{moddecay}.  This reduces the branching ratios for the
standard search signatures without significantly increasing the
production cross section, since gluon fusion dominates the production.
This might be one of the most difficult channels to uncover.

Alternatively we can again consider decays to $\phi$s which promptly
decay and back into SM quarks or gluons.  This hides the Higgs not
only by decaying into jets but these jets are also softer.  Examples
of these scenarios are realized in
Refs.~\cite{Chang:2008cw,Bellazzini:2009xt,Falkowski:2010hi}.
Introducing cascades as in Eq.~\eqref{eq:cascades} produces more but
softer radiation.  Therefore to identify the Higgs decay products it
is advantageous to produce the Higgs in a boosted state.  Irrespective
of the presence of a cascade this can help disentangle the Higgs from
the background ~\cite{Falkowski:2010hi,subjethiggs}.  This also helps
when one uses a cascade ending in partially or completely leptonic
final states which would otherwise be too soft to pass the
cuts~\cite{Cheung:2009su}.

Another option is to decay the Higgs into gluons via a $\phi$ with a
mass close to a heavy flavor bound state, e.g. $b\bar{b}$.  In the
$\phi$ rest frame the quarks are back to back and if their momenta are
small it is likely for them to hadronize to a quark-anti quark bound
state. Quark-anti quark bound states like $\Upsilon$ and $\eta_{b}$
typically decay dominantly into gluons.  In this case no direct
coupling of the $\phi$ to gluons and light quarks is necessary and
this would be a generic feature if the $\phi$ couples Yukawa-like to
quarks.  One example is a cascade $h\to \phi\phi\to
\Upsilon\Upsilon\to 6 g$.  In Fig.~\ref{fig:buried} we use
{\sc{Pythia}} 8 \cite{pythia8} to simulate such a cascade.  This
example demonstrates that the $\phi$ needs to be quite close to
threshold as already a relatively small mass difference between the
$\phi$ and the quark-anti quark bound state significantly reduces the
branching ratio to this bound state leaving us with a significant
number of long lived $B$-hadrons.  For other ``heavy'' quarks
identification is more difficult to begin with as there are fewer
clean long-lived states.

\subsection{Combined Higgs hide-outs}
\label{subsec:3}
The hide-outs outlined in the previous sections
\ref{subsec:1}-\ref{subsec:2} can be combined and result in a large
variety of signatures some of which can be even more challenging.

Let us imagine that the Higgs can decay via three different states
denoted by $\phi_{\text{inv}}$, $\phi_{\text{disp}}$, and
$\phi_{\text{jets}}$, where the subscript indicates the decay
properties of the $\phi$ as invisible, long-lived or dominant decay
into jets.

\bigskip

$h\to\phi_{\text{disp}}(\phi_{\text{inv}},\phi_{\text {jets}})$:
Adapted trigger strategies to highly displaced vertices as outlined in
\cite{trigger} can obviously cope with signatures of these types as
well, unless the majority of decays ocur outside the
detector. Eventually reconstructing the mass of the Higgs boson is
however more involved. This is due to the systematic uncertainties
that enter in the measurement of the $\slashed{E}_T$, which is
sensitive to all energy deposited in the detector, or the jet energy
scale.

\bigskip

$h\to\phi_{\text{inv}}\phi_{\text {jets}}$: If the Higgs is produced
at rest and $m_{h}\gg m_{\phi_{\text{jets}}}$ the resulting signature
is a monojet with a missing energy $\sim m_{h}/2$. In this case
boosting the Higgs by recoiling it against a $Z$ or an additional jet
is not necessarily a good strategy. Then the jet from
$\phi_{\text{jets}}$ and the missing energy from $\phi_{\text{inv}}$
are aligned.  This is problematic since it is already difficult to
precisely measure the jet energy. Therefore it is difficult to measure
missing energy aligned with jets. In fact, {\sc{Atlas}} and {\sc{Cms}}
typically apply cuts on extra jets in $\slashed{E}_T$ searches to
remove backgrounds, reducing the sensitivity toward these final
states~\cite{exisusy}. $Z+$jets is already a non-negligible background
if the Higgs recoils against a $Z$ but this problem becomes even more
severe if the Higgs recoils against a jet.

\section{Summary}
\label{sec:summary}
In this paper we have discussed phenomenological modifications of the
Higgs sector which can serve to hide a Higgs and weaken the
currently existing bounds.  The Standard Model-like Higgs is a minimal
solution to electroweak symmetry breaking.  Therefore we focus on
scenarios where the electroweak symmetry breaking sector is as in the
Standard Model (SM).  This limits the number of potential hide-outs
for the Higgs boson and allows to classify them in terms of
phenomenological signatures at colliders. We also suggest a simple
benchmark model that parameterizes these possible signatures.
Modified search strategies, looking at different signatures but also
using adapted trigger strategies and elaborate reconstruction
techniques (e.g. subjets) can help to uncover even an evasive Higgs.

There are essentially three possible hide-outs. One is to have
dominantly invisible decays. This typically produces missing energy
signatures.  In this case a promising search channel is a monojet plus
missing energy. We have used existing {\sc{Atlas}} data for this
channel and constrained the production cross section times invisible
branching fraction. For comparison we also estimated the sensitivity
for searches using associated production.

A second hide out is for the Higgs is to first decay into long-lived
neutral particles. Their eventual decay may then be observed in
displaced vertex searches. Depending on the decay length it may be
advantageous to also search for such decays occurring in the outer
parts of the detector.  This feature could even be enhanced by cascade
decays.

Finally, the Higgs could decay dominantly to light quarks and gluons
and be buried in huge QCD background.  These decays can occur via
cascades potentially including also intermediate $\bar{q}q$ bound
states.  Here, methods looking into specific radiation patterns in the
jet substructure can be important to uncover the Higgs.

Of course, combining the different hide-outs can make searches even
more difficult. 

In general we find that hiding heavy Higgses is rather challenging
from a theoretical point of view while light Higgses (for which $h\to
VV$ is inaccessible) is more straightforward.

Current analysis are not yet sensitive enough to pin down all Higgs
hide-outs.  However, the ever increasing LHC data set together with a
combined analysis of all sensitive channels will significantly tighten
the constraints for all hide-outs in the near future.

\section*{Acknowledgements}
We thank Kristov Hackstein, Marc Hohlfeld, Peter Loch, Andrew
Pilkington and David Strom for helpful discussions.  CE acknowledges
funding by the Durham International Junior Research Fellowship scheme.
ER acknowledges financial support from the LHCPhenoNet network under
the Grant Agreement PITN-GA-2010-264564.

The simulations underlying this study have been performed in parts on
bwGRiD, member of the German D-Grid initiative, funded by the
Bundesministerium f\"ur Bildung und Forschung and the Ministerium
f\"ur Wissenschaft, Forschung und Kunst Baden-W\"urttemberg.

\end{document}